%% file: main.tex
\documentclass[a4paper,fleqn]{cas-sc}

\usepackage[numbers]{natbib}
\usepackage{amsmath}
\usepackage{bm} 
 \usepackage{subcaption}
\def\tsc#1{\csdef{#1}{\textsc{\lowercase{#1}}\xspace}}
\tsc{WGM}
\tsc{QE}
\tsc{EP}
\tsc{PMS}
\tsc{BEC}
\tsc{DE}

\begin{document}

\title [mode = title]{Concurrent n-scale modeling for non-orthogonal woven composite}                      



\author[1]{Jiaying Gao}
\credit{Conceptualization, Methodology, Software, Validation, Formal analysis, Investigation, Data curation, Visualization, Writing - original draft}
\author[2]{Satyajit Mojumder}
\credit{Methodology, Software, Formal analysis,  Data curation, Visualization, Writing - original draft, Writing - review \& editing}
\author[1,3]{Weizhao Zhang}
\credit{Experiment,  Writing - original draft}
\author[1]{Hengyang Li}
\credit{Software,  Writing-Review}
\author[1]{Derick Suarez}
\credit{Software, Formal analysis,  Data curation, Writing - original draft, Writing - review \& editing}
\author[1]{Chunwang He}
\credit{Software, Data Curation}
\author[1]{Jian Cao}
\credit{ Writing - review \& editing, Supervision, Funding Acquisition}
\author[1]{Wing Kam Liu} {\corref{cor1}}
\credit{Conceptualization, Methodology,  Writing - review \& editing, Supervision, Funding Acquisition}
\cortext[cor1]{Corresponding author: Wing Kam Liu, Email: w-liu@northwestern.edu}

\address[1]{Department of Mechanical Engineering, Northwestern University, 2145 Sheridan Road, Tech B224, Evanston, IL 60208-3109, USA}
\address[2]{Theoretical and Applied Mechanics Program, Northwestern University, 2145 Sheridan Road, Tech B224, Evanston, IL 60208-3109, USA}
\address[3]{Currently at Department of Mechanical and Automation Engineering, The Chinese University of Hong Kong, Shatin, N.T., Hong Kong SAR, China}

\begin{abstract}
Concurrent analysis of composite materials can provide the interaction among scales for better composite design, analysis, and performance prediction. A data-driven concurrent n-scale modeling theory ($\textrm{FExSCA}^\textrm{n-1}$) is proposed in this paper utilizing a mechanistic reduced order model (ROM) called self-consistent clustering analysis (SCA). We demonstrated this theory with a  $\textrm{FExSCA}^2$ approach  to study the  3-scale woven carbon fiber reinforced polymer (CFRP) laminate structure. $\textrm{FExSCA}^2$ significantly reduced expensive 3D nested composite representative volume elements (RVEs) computation for woven and unidirectional (UD) composite structures by developing a material database. The modeling procedure is established by integrating the material database into a woven CFRP structural numerical model, formulating a concurrent 3-scale modeling framework. This framework provides an accurate prediction for the structural performance (e.g., nonlinear structural behavior under tensile load), as well as the woven and UD physics field evolution. The concurrent modeling results are validated against physical tests that link structural performance to the basic material microstructures. The proposed methodology provides a comprehensive predictive modeling procedure applicable to general composite materials aiming to reduce laborious experiments needed.
\end{abstract}

\begin{graphicalabstract}
	\includegraphics[width=1\textwidth]{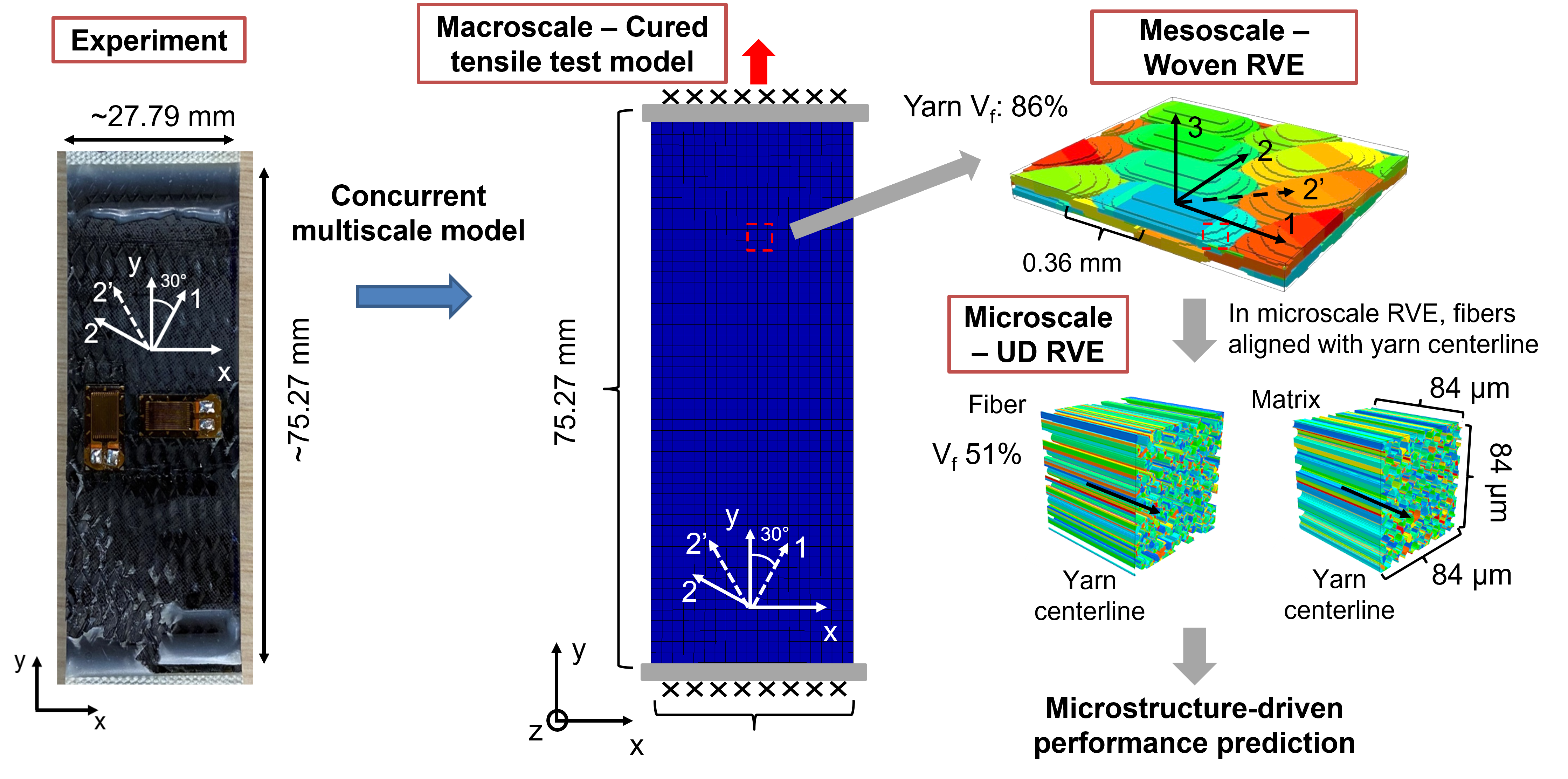}
\end{graphicalabstract}

\begin{highlights}
\item Theory for n-scale concurrent modeling of composite materials is proposed 
\item A 3-scale concurrent model accurately predicts non-orthogonal woven composite tensile experiment
\item Concurrent model is predictive of localized field evolution and nonlinear materials physics
\end{highlights}


\maketitle

\section{Introduction}
Numerical methods have been developed in the past few decades to facilitate modeling and simulations of engineering materials systems. For example, Finite Element Analysis (FEA) for vehicle crash simulation has become a standard procedure in major vehicle manufacturers to virtually examine vehicle safety under various scenarios. Numerical models can reduce physical experiments needed and provide significant savings in time and resources during the product development process. However, materials physics linked into multiple length scales need to be considered to properly model the materials system through any numerical procedure. Concurrent multiscale modeling approach provides the interaction among materials length scales that is crucial to incorporate materials physics and build a thorough understanding of the materials system.

Composite materials are in general composed of at least two different phases. For example, the cured unidirectional (UD) carbon fiber reinforced polymer (CFRP) is made of continuous carbon fibers and polymer matrix. Prediction of UD CFRP properties can be carried out using a UD microstructure model (usually a Representative Volume Element (RVE)) for numerical homogenization that is performed through FE or Fast Fourier Transformation (FFT) methods. Various UD CFRP properties, the elastic, elasto-plastic, as well as the damage behavior of composites can be predicted by the RVE model \cite{Kari2007,Bradshaw2003,Liang2019,gao2017multiscale,Sun2018,Sun2019}. RVE analysis provides a good estimation of composite material constants, and it can be used for hierarchical multiscale modeling where the RVE output serves as the material law for the structural level model \cite{li2019clustering, he2020hierarchical}. Such an approach applies to woven composites as well, as seen in \cite{Liu2016, Dasgupta1996, liang2019multi}. In the macroscale woven laminate, its material properties can be approximated with mesoscale woven RVEs, and in each woven RVE, the yarn phase material properties can be approximated with microscale UD RVEs as each yarn is a mixture of many carbon fibers and the epoxy matrix. Building a hierarchical multiscale model would allow one to incorporate mesostructure and microstructure elastic constants into the woven laminate \cite{bostanabad2018uncertainty,he2019multiscale,he2020hierarchical}, and hence the laminate performance can be evaluated with homogenized information, yet such approach implies simplification on the nonlinear material behaviors.

For accurate prediction of engineering structures, one needs to establish the capability of efficient computation of CFRP nonlinear material responses. This requires the development of reduced-order modeling methods so the RVE model can be effectively embedded into the structural level model \cite{liu2016self, liu2018microstructural,yuan2008toward, wulfinghoff2018model}. Composite concurrent multiscale modeling is established based on a reduced order modeling approach. Under the concurrent modeling framework, a RVE is replaced by a compressed RVE database, namely the reduced order model, to replace traditional material law approach in composite structural performance prediction. Two-scale (2-scale) concurrent modeling for UD and woven composites has been developed utilizing Self-consistent Cluster Analysis (SCA), a reduced order modeling method, as reported in \cite{han2020efficient, gao2020predictive}. However, for 2-scale woven multiscale modeling, yarn properties are usually assumed to be linear elastic \cite{bostanabad2018uncertainty, gao2017multiscale, han2020efficient}. In this paper, a three-scale (3-scale) concurrent multiscale modeling approach, $\textrm{FExSCA}^2$, is proposed in order to properly capture yarn plasticity (which is approximated using UD RVEs) during the woven structure deformation process. The outcomes are two-folded: (1) the prediction accuracy is improved by incorporating previously ignored physics, and (2) the history of the plastic strain accumulation in woven and UD scale is recorded for a better understanding of microstructure evolution.

The outline of the paper is as follows: Section 2 provides materials and methods  for experimental woven bias-extension sample test. Section 3 provides a concurrent n-scale modeling theory  for heterogeneous materials using a mechanistic reduced order model. A validation of the proposed concurrent modeling theory is presented in Section 4. Section 5 describes the concurrent multiscale modeling setup for cured woven tensile sample. Section 6 presents key simulation results, findings, and comparisons with experimental data. Section 7 concludes the paper with possible future directions.

\input{test}
\input{concurrent_theory}

\input{Validation_bending}
\input{modeling_bias}

\input{result}
\section{Conclusion}
In this paper, a three-scale concurrent multiscale modeling theory is presented with a demonstration of 3-pt bending woven composite laminate and cured woven bias extension test. The proposed concurrent multiscale theory captured the extra plasticity in the mesoscale woven RVE yarn phase through modeling the yarn with microscale UD RVEs. Without accounting for the extra plastic deformation in the yarn phase, the force-axial strain curve predicted by the 2-scale model diverges far away from the test data. The 3-scale model, on the other hand, is able to capture the correct trend of the loading force history with a reasonable accuracy. On top of that, by modeling the yarn phase with UD at different fiber volume fractions, the upper and lower bounds of the force-strain curve can be captured. This makes the 3-scale model a powerful tool for modeling woven composite structure with the consideration of the fabrication tolerance in the UD fiber volume fraction. Such capability can help building a better understanding of the woven structural performance, accelerating the search for the optimal composite fiber volume fraction in the woven composite structure by avoiding excessive experimental trial and error. In the future, failure criteria for the 3-scale models can be added to expand the capability of the concurrent modeling theory in predicting failure behavior.

\appendix
\section{Formulation of the reduced order modeling method}\label{app.A}
The reduced order modeling method utilized the incremental form of the Lippmann-Schwinger equation given in Eq. \ref{eq:LP_inc}, of which the FFT method was based on.
\begin{eqnarray} \label{eq:LP_inc}
\Delta\bm{\varepsilon(\bm{x})}=\Delta\bm{\varepsilon}^{M}-\int_{\Omega}\bm{\Gamma^{0}}(\bm{x},\bm{x}'):(\Delta\bm{\sigma(x')}-\bm{C}^{0}:\Delta\bm{\varepsilon(x')})d\bm{x'}
\end{eqnarray}
where $\Delta\bm{\varepsilon(\bm{x})}$ is the RVE voxel-wise incremental strain tensor, $\Delta\bm{\sigma(x)}$ is the RVE voxel-wise incremental stress tensor, $\bm{C}^{0}$ is the stiffness tensor of the elastic reference material, and $\bm{\Gamma}^{0}$ is the isotropic Green's function based on reference material $\bm{C}^{0}$, x represents integration point within each cluster. $\Delta\bm{\varepsilon}^{M}$ is the strain increment passed in from the macroscale calculation. Note that the stress tensor at each integration point can be computed using elastic and elasto-platic material laws. In this work, isotropic elasticity, transversely isotropic elasticity, and J2 plasticity are used to evaluate the stress tensor at each integration point. 

For each RVE, assuming its voxels can be classified into a limited number of clusters, where each cluster share the same strain and stress increment, it is possible to arrive at the following discretized Lippmann-Schwinger equation in Eq. \ref{eq:LP_dis}
\begin{eqnarray} \label{eq:LP_dis}
\frac{1}{c^{I}|\Omega|}\int_{\Omega}\chi^{I}(\bm{x})\Delta\bm{\varepsilon(\bm{x})}d\bm{x}=\Delta\bm{\varepsilon}^{M} -\frac{1}{c^{I}|\Omega|}\int_{\Omega}\int_{\Omega}&\chi^{I}(\bm{x})\bm{\Gamma^{0}}(\bm{x},\bm{x}'):(\Delta\bm{\sigma(x')}-\bm{C}^{0}:\Delta\bm{\varepsilon(x')})d\bm{x'}d\bm{x}
\end{eqnarray}
where $|\Omega|^{I}$ is the domain volume of cluster $I$, $c^{I}$ is the volume fraction of cluster $I$ in the RVE  domain, and $|\Omega|$ is the RVE domain volume. The $\chi(x)$ function equals 1 when $x$ is in the cluster $I$, otherwise it will be 0. This allows one to further simplify the discretized Lippmann-Schwinger equation to Eq. \ref{eq:LP_dis_final}
\begin{equation} \label{eq:LP_dis_final}
\Delta\bm{\varepsilon}^{I}=\Delta\bm{\varepsilon}^{M}-\sum_{J=1}^{K}\left[ \frac{1}{c^{I}|\Omega|}\int_{\Omega}\int_{\Omega}\chi^{I}(\bm{x})\chi^{J}(\bm{x'})\bm{\Gamma^{0}}(\bm{x},\bm{x}')d\bm{x'}d\bm{x}:(\Delta\bm{\sigma}^{J}-\bm{C}^{0}:\Delta\bm{\varepsilon}^{J})\right]
\end{equation}

One should notice that in Eq. \ref{eq:LP_dis_final}, the Green's function can be separated out, forming a constant interaction tensor for the ROM. The interaction tensor is defined in Eq. \ref{eq:DIJ}
\begin{equation} \label{eq:DIJ}
\bm{D}^{IJ}=\frac{1}{c^{I}|\Omega|}\int_{\Omega}\int_{\Omega}\chi^{I}(\bm{x})\chi^{J}(\bm{x'})\bm{\Gamma^{0}}(\bm{x},\bm{x}')d\bm{x'}d\bm{x}
\end{equation}

Finally, the discretized Lippmann-Schwinger equation can be written as Eq. \ref{eq:LP_dis_fina_Dij}, which can be solved with Newton's method. Comparing Eq. \ref{eq:LP_dis_fina_Dij} to Eq. \ref{eq:LP_inc}, only $K$ strain tensors needs to be evaluated, and $K$ is much less than the number of voxels in the original RVE. Therefore, the ROM can significantly reduced the overall computational cost of the 3-scale concurrent model, while allowing one to track the RVE strain and stress evolution. 
\begin{equation} \label{eq:LP_dis_fina_Dij}
\Delta\bm{\varepsilon}^{I}=\Delta\bm{\varepsilon}^{M}-\sum_{J=1}^{K}\left[ \bm{D}^{IJ}:(\Delta\bm{\sigma}^{J}-\bm{C}^{0}:\Delta\bm{\varepsilon}^{J})\right]
\end{equation}

\section{ Concurrent stress evolution in three-point bending test of woven laminate} \label{app.B}
\begin{figure}[h]
	\centering
	\begin{subfigure}[b]{.45\linewidth}
	   		\centering
		\includegraphics[width=\linewidth]{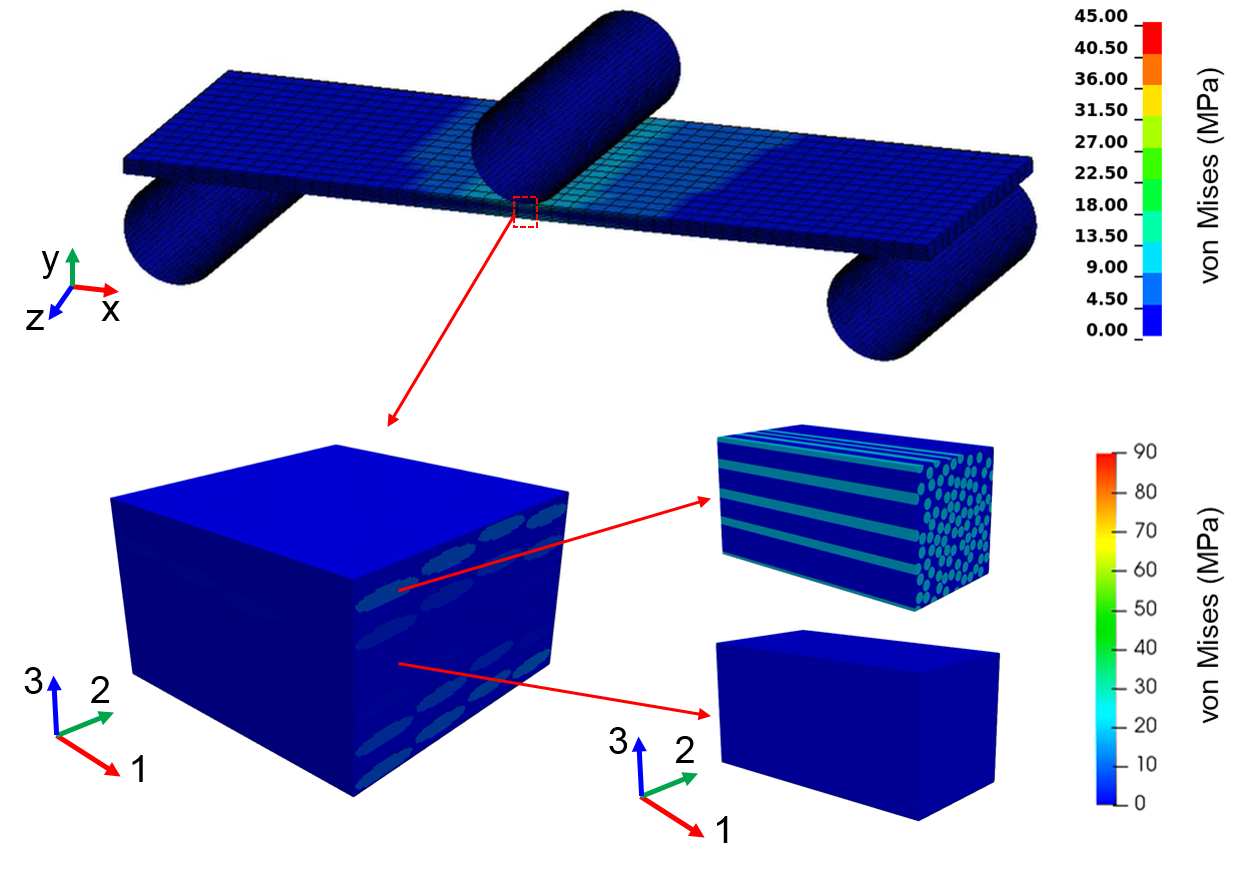}
		\caption{Impactor displacement of 0.16 mm} 
	\end{subfigure}
	\begin{subfigure}[b]{.45\linewidth}
	   		\centering
		\includegraphics[width=\linewidth]{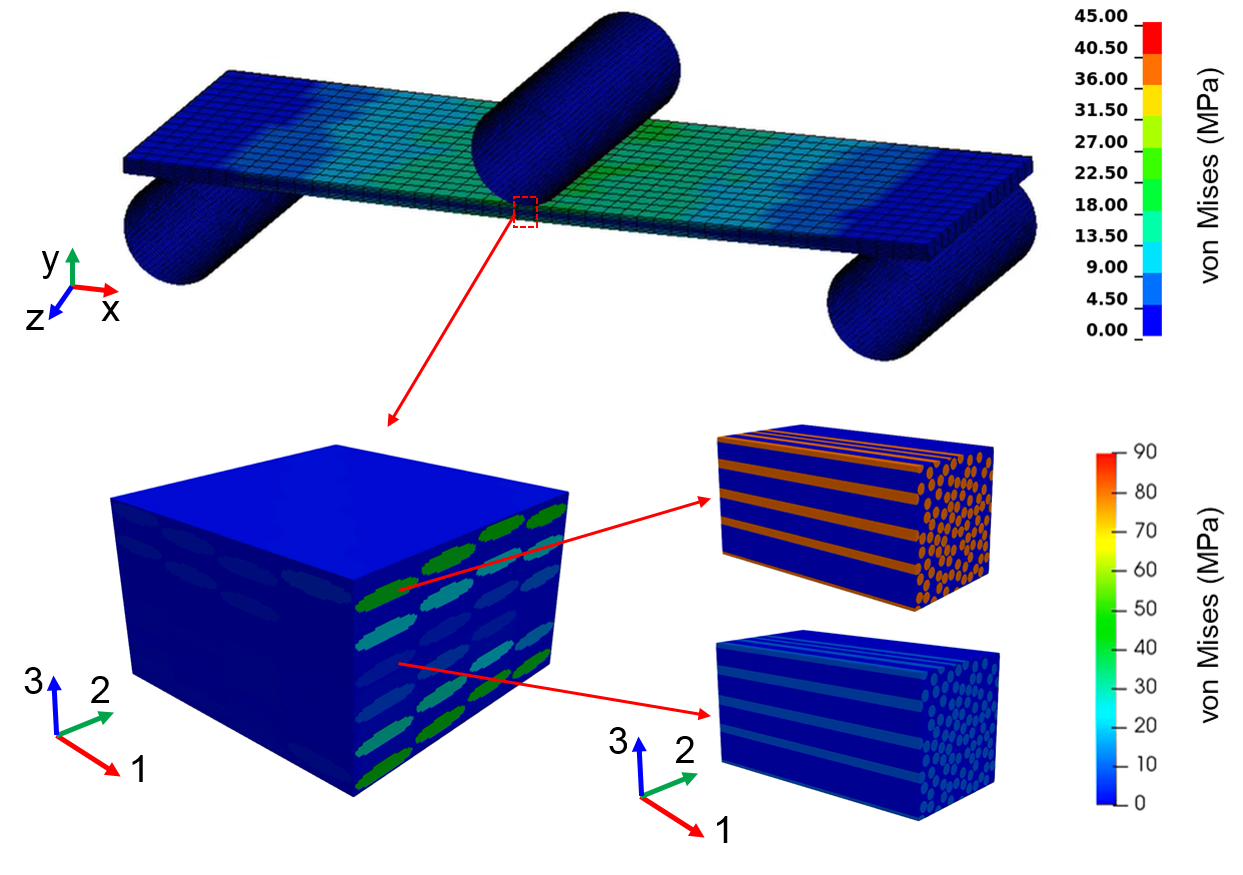}
		\caption{Impactor displacement of 0.32 mm}
	\end{subfigure}
	\begin{subfigure}[b]{.45\linewidth}
	   		\centering
		\includegraphics[width=\linewidth]{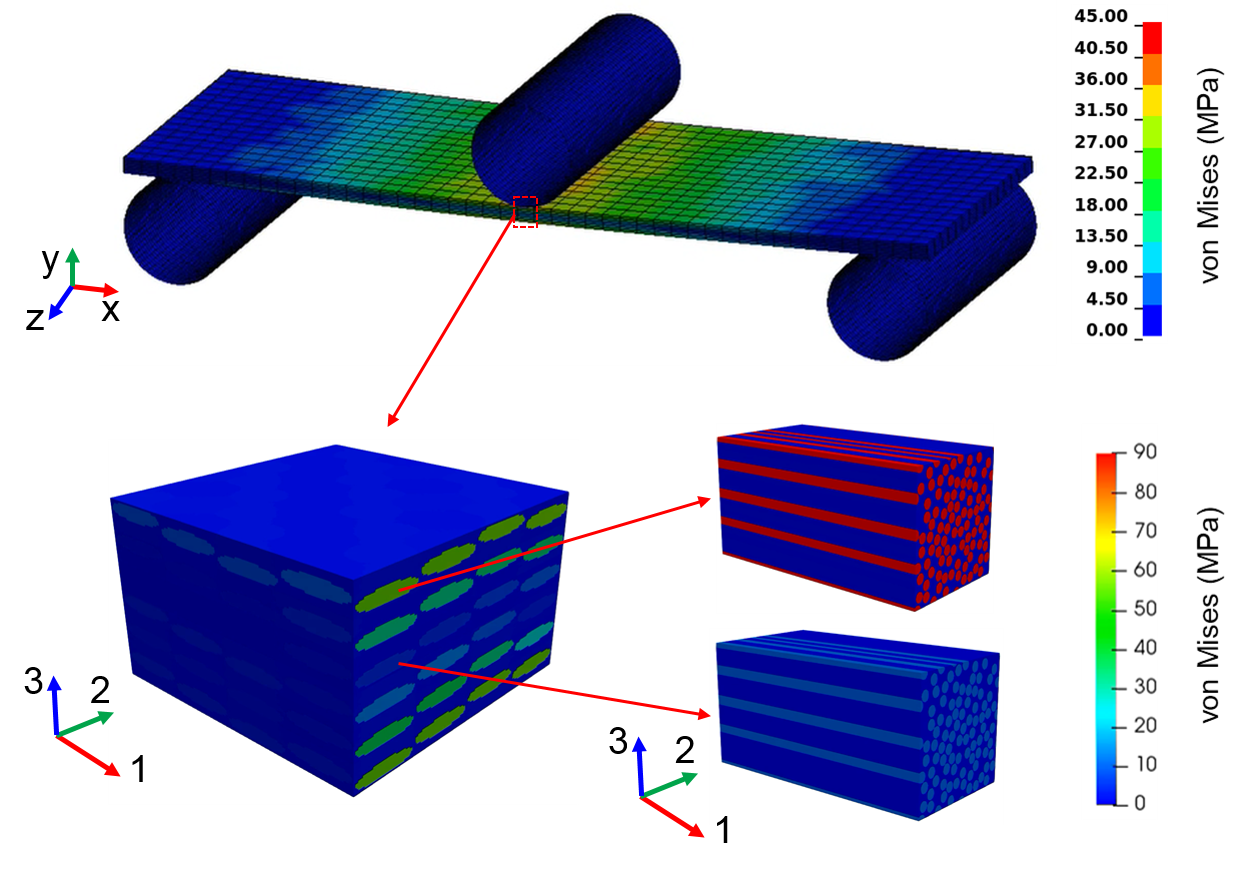}
		\caption{Impactor displacement of 0.48 mm}
	\end{subfigure}
	\caption{Three snapshots to illustrate the macroscale (woven laminate), mesocale (woven RVEs) and microscale (UD RVEs) stress evolution.}
	\label{fig:3scale_snapshots}
\end{figure}

In Figure \ref{fig:3scale_snapshots} a) to c) the von Mises stress fields evolution for macroscale woven laminate, mesoscale woven microstructures, and microscale UD microstructures are visualized. Figure \ref{fig:3scale_snapshots} a), b), and c) are taken at impactor displacement of 0.16 mm, 0.32 mm, and 0.48 mm, respectively. Note that woven RVEs for all 6 plies, representing the outermost region in the middle of the woven laminate, are plotted. The two UD RVEs depicted represent two yarns, one in the first ply and one in the third ply.

As the impactor gradually bends the laminate, the von Mises stress magnitude of the laminate increases. To be specific, the top and bottom layer are experiencing compression and tension stresses, respectively. The neutral surface of the laminate, which is between the third and fourth ply, shows very little increase in the von Mises stress, as expected. This is illustrated by the woven microstructure plot depicted using six RVEs extracted from six plies. Figure \ref{fig:3scale_snapshots} shows the von Mises stress of the top and bottom layer increasing rapidly, whereas plies in the middle experience a very small increase. For further illustration, the yarn microstructure responses are captured by UD RVEs. The fiber phase in the UD RVE representing the yarn in the first woven ply experiences a much higher increase in the von Mises stress magnitude, whereas the fiber phase in the third ply bares much less stress. The laminate bending causes bending stress along the x direction of the laminate, which coincides with the 1 direction in the yarn. Therefore, one can see at the mesoscale, yarns in the 1 direction are the primary load-carrying constituents. As the yarn responses are approximated by the UD RVE, and the UD fiber direction coincides with the woven 1 direction, fiber phase in the UD RVE carries the load, as expected. Therefore, aforementioned evolution of the von Mises stress across all three scales is physical, and it reveals underneath microstructure evolution in the woven laminate structure only enabled by the 3-scale concurrent modeling.

\printcredits

\section*{Acknowledgements}
The financial support for this project was provided by AFOSR (FA9550-18-1-0381) and National Science Foundation’s Mechanics of Materials and Structures (MOMS) program under the Grant No. MOMS/CMMI-1762035. 

\bibliographystyle{model1-num-names}

\bibliography{cas-refs}

\end{document}

%% file: test.tex
\section{Materials and methods}
\label{sec:woven_bias}
\subsection{Materials and sample geometry}
\begin{figure}[!htb]\noindent
  \centering\includegraphics[width=0.3\linewidth]{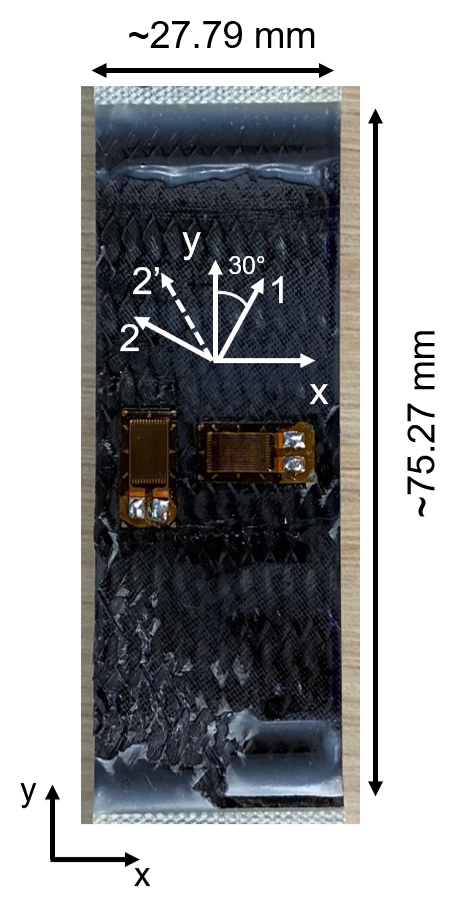}
  \caption{The cured woven bias-extension sample (clamping region hidden) after testing, as seen by broken epoxy matrix and exposed yarns. 1 and 2 represent weft and warp yarns for the orthogonal configuration, 90$^{\circ}$ yarn angle. The sample has a tailored yarn angle of 60$^{\circ}$ represented by 1 and 2$^{'}$.}%
  \label{fig:60_woven}
\end{figure}

In this work, bias extension tests are performed on cured woven prepregs with 60$^{\circ}$ yarn angle. First, woven prepregs with 90$^{\circ}$ yarn angle from Dow Chemical are used to manufacture the 60$^{\circ}$ non-orthogonal samples. Bias extension tests are performed on the orthogonal prepregs to  generate shear deformation in order to achieve a 60$^{\circ}$ yarn angle. The bias-extension tests are performed at elevated temperature to reduce the resistance of uncured resin during shear deformation. The woven prepregs are trimmed and have an initial length that is at least twice the sample width. The prepregs with 60$^{\circ}$ yarn angles are then cured to make tensile test samples. The cured woven samples have overall fiber volume fractions between 44$\%$ and 48$\%$ per measurements done according to ASTM standard D792-01\cite{standardd792}. The 60$^{\circ}$ yarn angle and the shape of the cured woven sample are visualized in Figure \ref{fig:60_woven}. Detailed information on performing bias-extension tests and the curing process can be found in \cite{liang2019multi}.

\subsection{Experimental setup}
Tensile tests are performed using a hydraulic testing machine with a load cell capacity of 100 kN. The cured woven samples are loaded on the testing machine, clamped on top and bottom ends, and deformed along the y axis denoted in Figure \ref{fig:60_woven}. Axial strain is recorded by the strain gauge, and the axial loading force is recorded by the load cell. The data gathered from the tensile tests are used to generate the force vs. axial strain plots that are used for validation purposes. The tensile test is performed on three samples to ensure repeatability and quality of the force-strain curves. 

\subsection{Computational model development}
\begin{figure}[h]
	\centering
	\includegraphics[width=0.9\textwidth]{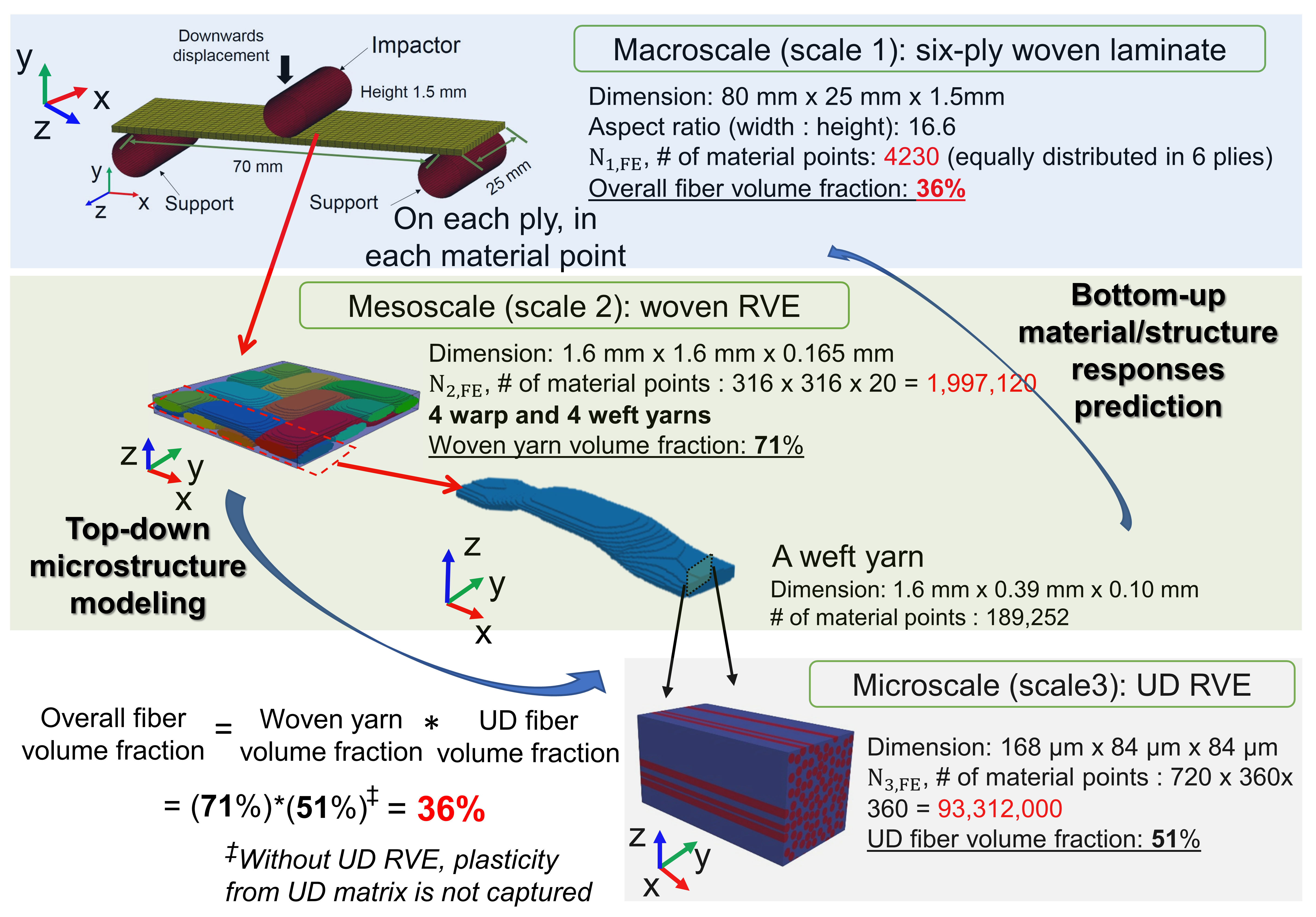}
	\caption{LS-DYNA macroscale finite element model of a woven laminate under 3-point bending is shown on the top panel. Each material point in the macroscale laminate is represented by a mesoscale woven RVE shown in the middle panel and a material point in the yarns (weft yarn shown) is represented by a microscale UD RVE shown in the bottom panel. An estimate of the overall fiber volume fraction captured through these three scales is shown.}
	\label{fig:woven_3pt_bending}
\end{figure}

The woven composite structure is a good demonstration of the length scales of the composite materials. To model a woven composite structure, we first took a 3-point bending woven composite laminate and attempt to develop a computational framework that can be applicable to the woven tensile sample.  The computational complexity of 3-point bending of a woven composite laminate is shown in Figure \ref{fig:woven_3pt_bending}. Woven composite laminate can be modeled at part scale with multiple plies using finite element analysis. Each material point of the laminate can be represented by a mesoscale woven RVE (4x4 twill here) having multiple yarns. A zoomed in view of the yarn reveals a UD structure at the microscale. To fully model the laminate concurrently, one would need to solve approximately $10^{18}$ degrees of freedom (as shown in the figure), which is not computationally feasible. Therefore, an efficient multiscale theory is required to model multiple scales together. Utilizing the data-driven techniques of mechanistic reduce order modeling, this problem becomes tractable with a desktop computer. This type of multiscale model is necessary to properly capture the materials physics at multiple length scales. Solely using a 2-scale model for the woven composite laminate is not able to consider the plasticity from the UD composite. Moreover, the overall fiber volume fractions can be captured more accurately if more scales are considered.  

%% file: concurrent_theory.tex
\section{Concurrent theory for n-scale materials modeling}

\subsection{General n-scale theory}

\begin{figure}
	\centering
	\includegraphics[width=0.70\textwidth]{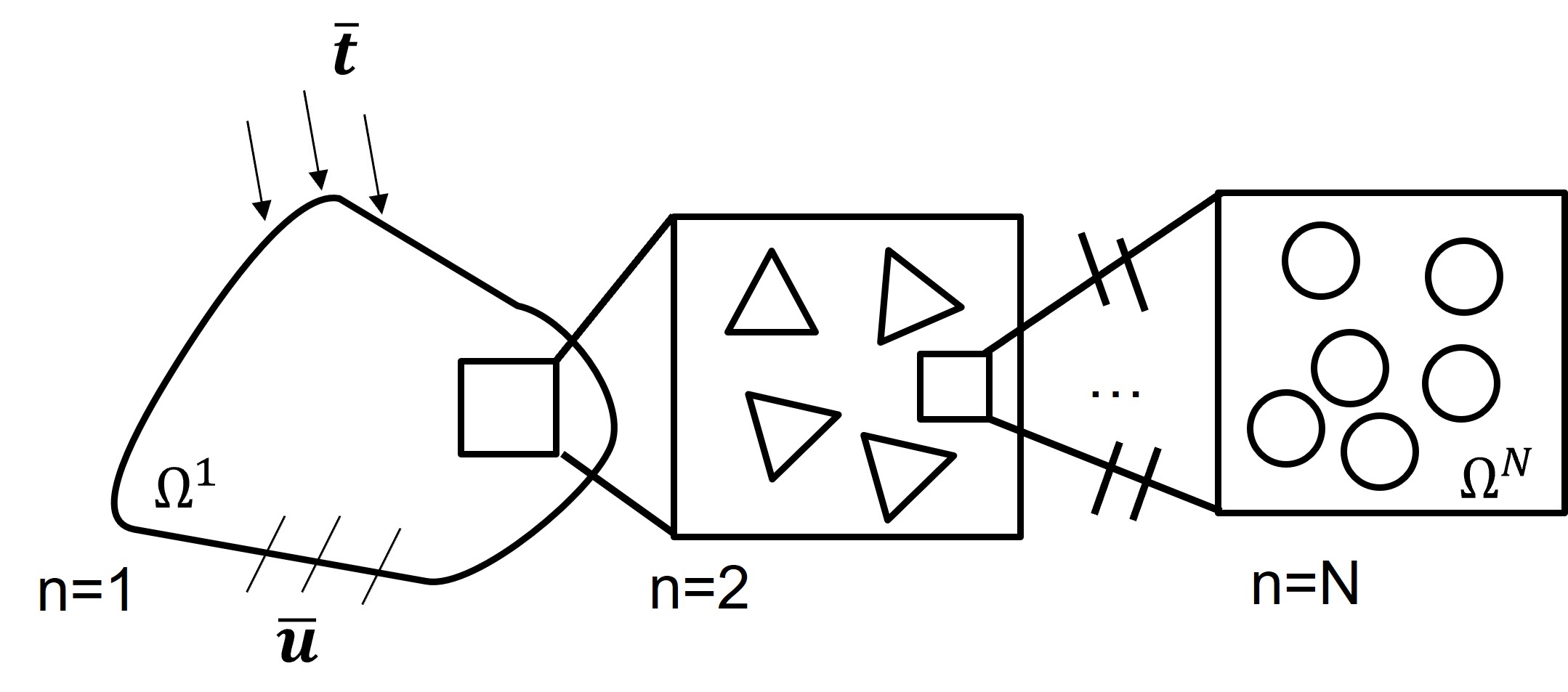}
	\caption{A general illustration of the N-scale multiscale problem. n $=$ 1 is the macroscale, and n $>$ 1 are finer scales representing the underneath heterogeneity \cite{yu2021multiresolution}}
	\label{fig:multiscale_Nscale}
\end{figure}
Composites are heterogeneous materials showing multiple length scales. Figure \ref{fig:multiscale_Nscale} shows a composite material system with $n$ number of scales. Each scale has heterogeneity that can be further analyzed in a finer scale as shown in the figure. This general multiscale boundary value problem can be expressed using the equilibrium, compatibility equations with necessary boundary conditions and material law as given in Eq. \ref{eq:N_scale_conc} for n-scales. 

\begin{equation}\label{eq:N_scale_conc}
\begin{cases}
\nabla \cdot \bm{\sigma}^{1}+\bm{b}^{1}=0, \:\:\:in\:\Omega^{1}\\
\bm{t}=\bar{\bm{t}}, \:\:\:on\:\partial\Omega^{1}_{t}\\
\bm{u}=\bar{\bm{u}}, \:\:\:on\:\partial\Omega^{1}_{u}\\
\bm{\sigma}^{n}(\bm{X}^{n})=
\begin{cases}
\frac{1}{\Omega^{n+1}}\int_{\Omega^{n+1}}\bm{\sigma^{n+1}}(\bm{X}^{n+1})\Omega^{n+1}, if\:\bm{X}^{n}\:has\:microstructure,\:n = 2,...,N\\
f^{n}(\bm{\varepsilon}^{n})	, \:\:\:n = 1
\end{cases}\\
\nabla \cdot \bm{\sigma}^{n}=0, n = 2,...,N\\
\bm{\varepsilon}^{n}(\bm{X}^n)=\frac{1}{2}(\nabla\bm{u}^{n}(\bm{X}^n)+(\nabla\bm{u}^{n}(\bm{X}^n))^{T}), \forall \bm{X}^{n} \in \Omega^{n}, n = 1,2,3,...,N, \\
\end{cases}
\end{equation}
where $\bm{\sigma}^{1}$ is the macroscale stress tensor and $\bm{b}^{1}$ is the macroscale body force, $\bm{t}$ and $\bm{u}$ are the applied Dirichlet boundary conditions, and $\bm{\sigma}^{n}(\bm{X}^{n})$ and $\bm{\varepsilon}^{n}(\bm{X}^n)$ are stress and strain tensors at each lower scale $n$ integration point $\bm{X}^n$. 

In equation set \ref{eq:N_scale_conc}, the first three equations describe the macroscopic problem. The fourth equation defines the constitutive behavior for material points in scale $n$. If microstructure information needs to be captured, mean field homogenization is applied on scale $n+1$  to obtain effective stress responses for scale $n$. If microstructure is not needed, then the standard material constitutive law can be applied on $n$. For finer scales where $n>1$, microstructure responses are resolved assuming periodic displacement field and anti-periodic traction boundary condition, as described by the last two equations in Eq. \ref{eq:N_scale_conc}.

Assuming the case in which $N = 1$, the multiscale problem degenerates to a macroscopic problem described by continuum mechanics. Most modern engineering applications, such as vehicle crash simulations, are modeled at this scale. Finite Element Method (FEM) can be applied to solve the macroscale problem easily due to its generality in describing complex geometries \cite{jacob2007first,hughes2012finite,belytschko2013nonlinear}. Other methods, such as Isogeometric Analysis \cite{hughes2005isogeometric, cottrell2009isogeometric} and meshfree methods \cite{belytschko1994element,liu1995reproducing,li2002meshfree} are also applicable. However, a one-scale analysis means microstructure information cannot be directly captured, and the simulation process requires effective material laws \cite{goldberg2014theoretical}.  

\subsection{Reduced order model for concurrent analysis}

\begin{figure}[!htb]
	\centering
	\includegraphics[width=0.85\textwidth]{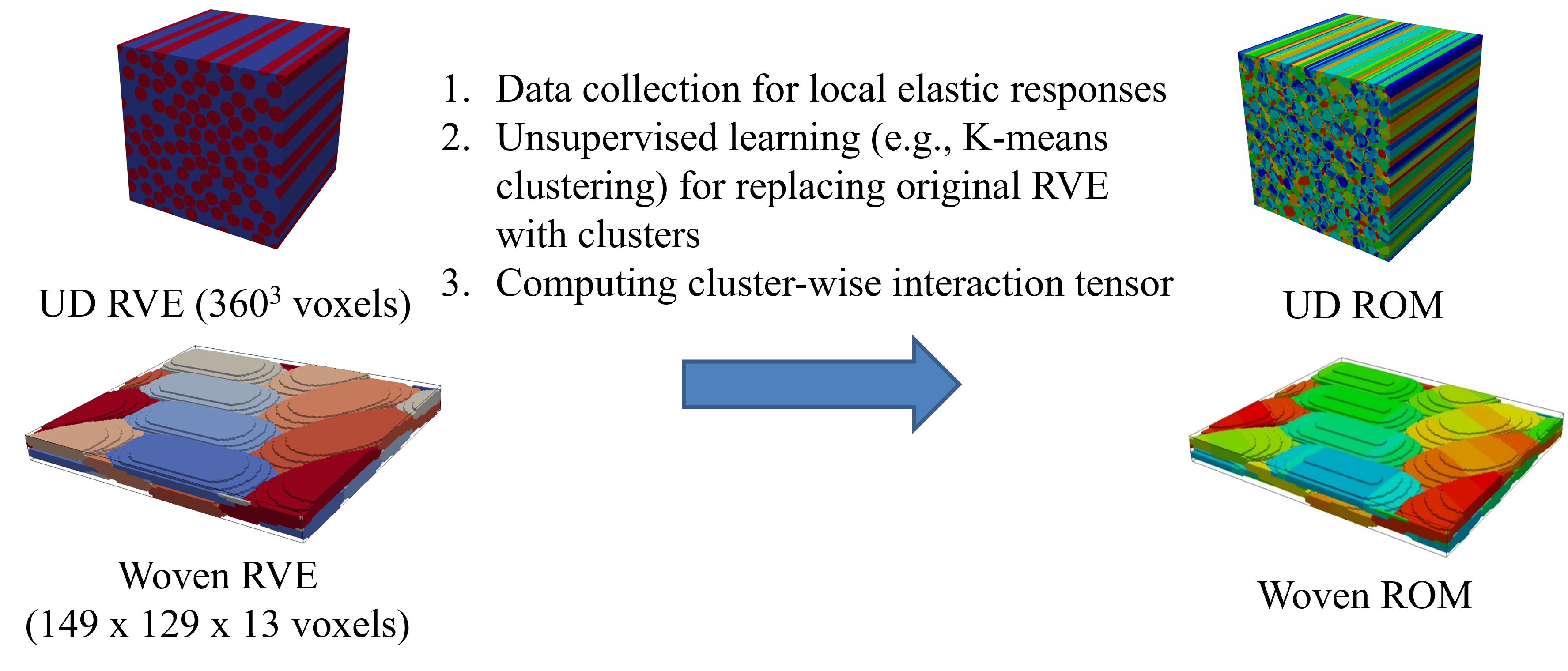}
	\caption{ROMs construction for UD and Woven RVEs, where voxel elements' boundary lines are hidden for clarity. Each RVE undergoes a three step data compression process (through unsupervised learning) in order to build the ROM. After the data compression process, the UD RVE is replaced with a ROM with 10 clusters (8 in the matrix phase and 2 in the fiber phase), and the woven RVE is replaced with a ROM with 40 clusters (8 in the matrix phase and 32 in the yarn phase). Same colored voxels belong to the same cluster within each ROM RVE.}
	\label{fig:RVE_to_ROM}
\end{figure}

From the general concurrent n-scale equations presented in Eq. \ref{eq:N_scale_conc}, each scale needs to be solved simultaneously. As shown in Figure \ref{fig:woven_3pt_bending}, considering three scales requires solving approximately $10^{18}$ degrees of freedom, which is not computationally feasible. At each lower scale, RVE responses are evaluated assuming periodic boundary conditions to ensure convergence by alleviating the size effect. Each mesoscale and microscale RVE are given an effective strain increment during the concurrent analysis. The RVE model then computes its effective RVE stress increment using a numerical homogenization technique, such as Fast Fourier Transformation (FFT). The effective stress increment is passed back to the higher scale in order to continue the concurrent analysis. However, direct RVE calculation is very expensive, as each RVE contains millions of voxel elements. To counteract this issue, a mechanistic reduced order modeling method, namely Self-consistent Clustering Analysis (SCA) \cite{liu2016self}, is leveraged. Details of the method are given in Appendix \ref{app.A}. Reduced-order models (ROMs) of RVEs are built to replace original high fidelity RVEs as shown in Figure \ref{fig:RVE_to_ROM}.

%% file: Validation_bending.tex
\section{Validation of computational model for 3-point bending test of woven laminate}
\label{sec:3pt-3-scale}
A common material testing procedure for determining the structural level properties under combined loading of composites laminates is the flexural test. The flexural test can be performed either as a 3-point bending test or 4-point bending test \cite{Khashaba2006a,Ullah2012c}. The flexural test can be used to investigate the elastic response of the composite laminate (such as flexural modulus), as well as the flexural strength if needed. To perform the virtual 3-point bending test, a macroscale woven laminate model is constructed. The 3-point bending test woven laminate structure is adopted from \cite{Ullah2012c}, depicted in Figure \ref{fig:woven_3pt_bending}. The laminate has a length of 80 mm (70 mm between two rigid supports), a width of 25 mm, and a thickness of 1.5 mm. The two circular supports have diameters of 10 mm, separated 70 mm apart from each other, as shown in the Macroscale part in Figure \ref{fig:woven_3pt_bending}. A loading nose or impactor of equal size as the two supports pushes the laminate downwards in the negative y direction. Carbon fiber and epoxy material properties used are given in Table \ref{table:woven_material_const}, though epoxy plasticity is not considered as the purpose of the test is to investigate the flexural modulus of the laminate structure.

\begin{table}[width=.4\linewidth,cols=6,pos=h]
 \caption{Elastic material constants for fiber (transversely isotropic elastic) and epoxy (isotropic elastic).} \label{table:woven_material_const}
\begin{tabular*}{\tblwidth}{c |c|c |c}
    \toprule 
    Fiber & Values & Matrix & Values\\
   \midrule
    $E_{11}$ & 245 GPa & $E$ & 3.8 GPa\\
    $E_{22}=E_{33}$ & 19.8 GPa & $\nu$ & 0.39 \\
    $G_{12}=G_{13}$ & 29.2 GPa &  \\
    $G_{23}$ &  5.92 GPa &  \\
    $\nu_{12}=\nu_{13}$ & 0.28 &  \\
    $\nu_{23}$ & 0.32 & \\
    \bottomrule
    \end{tabular*}
    \end{table}

\begin{figure}[h]
	\centering
	\includegraphics[width=0.9\textwidth]{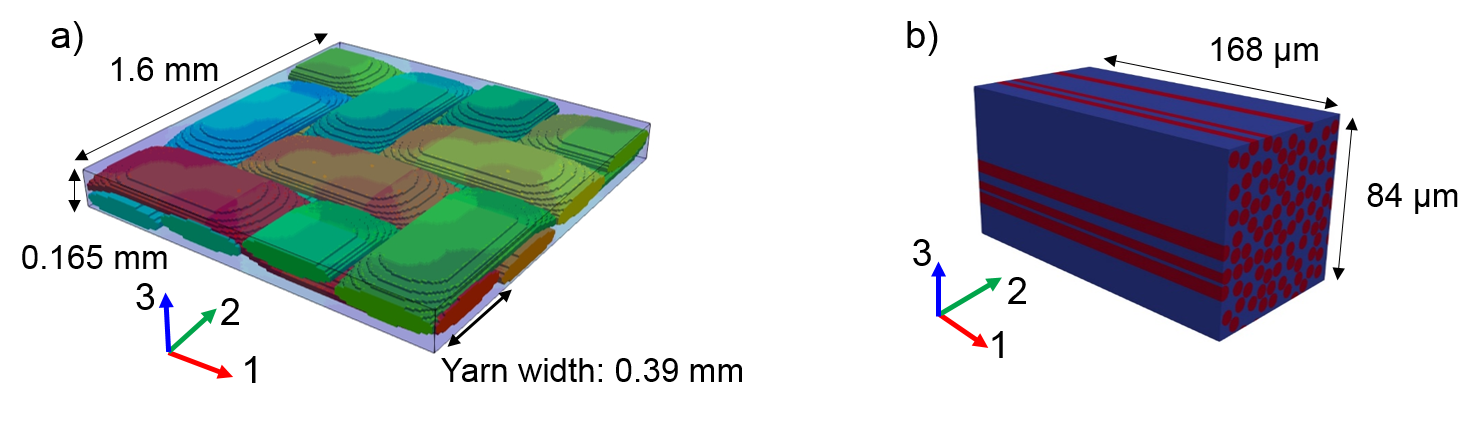}
	\caption{Geometric specification for a) woven RVE; b) UD RVE}
	\label{fig:woven_woven_ud_microstructure}
\end{figure}

In the concurrent model, the woven laminate is considered the macroscale. One material point of the macroscale laminate is modeled with a woven RVE at the mesoscale. The mesoscale woven RVE is depicted in Figure \ref{fig:woven_woven_ud_microstructure} a), with a  yarn width of 0.39 mm, length of 1.6 mm, and thickness of 0.165 mm. The woven yarn volume fraction is 71$\%$. The woven RVE containing 1,997,120 voxel elements is replaced by a ROM containing 4 clusters for the matrix phase and 16 clusters for the yarn phase. A fiber volume fraction of 51$\%$ is assumed for the yarn phase, and each cluster of the yarn phase is represented by a microscale UD RVE depicted in Figure \ref{fig:woven_woven_ud_microstructure} b). The UD RVE containing 93,312,000 elements is replaced by a ROM that has 4 clusters for the matrix phase and 2 clusters for the fiber phase.

The information passed in the 3-point bending concurrent modeling simulation is given in Figure \ref{fig:woven_3pt_bending}. Under this setup, at each integration point in the laminate model, a woven ROM is used to compute the material responses at that integration point. For all yarns in the woven ROM, UD ROMs are used to compute the yarn responses. The prediction of the flexural modulus is compared with the theoretical value, as shown in Table \ref{table:modulus_comp}. The flexural modulus follows the equation given below \cite{zweben1979test}:
\begin{equation}\label{eq:flexural_modulus}
E=\frac{L^{3}m}{4wh^{3}}
\end{equation}
where $m$ is the slope of the force-deflection curve computed from simulation, and $L$, $w$, and $h$ are length between two supports, width and height of the woven laminate. The theoretical value is computed from the woven RVE by applying loading along the local 1 (or 2) direction. Since the flexural modulus is one way to evaluate woven laminate's tensile behavior, a good match is observed.

\begin{table}[H]
	\caption{Comparison of theoretical (given by equation 2) and multiscale model prediction of flexural modulus}
	\label{table:modulus_comp}
	\centering
	\begin{tabular}{|c|c|c|c|} 
		\hline
		& Theoretical value & Multiscale model prediction  & Difference \\
		\hline
		Flexural modulus  & 48.77 GPa & 47.31 GPa  & 3.0$\%$ \\
		\hline
	\end{tabular}
\end{table}

The flexural modulus represents only the macroscopic property of the woven laminate. Another important feature of the 3-scale concurrent model is to provide microstructure physical fields (such as von Mises stress) evolution in the mesoscale and microscale, as shown in Appendix \ref{app.B}. A movie for the concurrent evolution of the stress field in the 3-point bending sample and the RVEs at different scales is provided as Supplementary material (S1). 

The 3-point bending example shows how 3-scale multiscale concurrent modeling can be used to predict both macroscopic flexural modulus and microstructure evolution. This establishes the baseline for conducting a virtual test for a woven laminate structure. For practical usage, realistic woven and UD microstuctures can be used as input to provide a microstructure driven prediction of structural performance. Alongside structural performance, microstructure evolution can be used to identify weak spots in both the mesoscale and microscale, and guide further design iterations for enhancing structural performance and integrity. Such virtual design process reduces number of experimental trials needed, and provides a trustworthy approach for obtaining the structural performance of composites.

%% file: modeling_bias.tex
\section{Concurrent modeling of cured woven tensile sample}
In  this section, the concurrent multiscale model development for the woven bias extension sample is described. The woven extension test sample is considered as a 3-scale problem where the macro scale is the sample itself, and meso- and micro-scale is the woven and UD composite scales, respectively. The macroscopic test sample itself is modeled with a layer of solid Finite Element (FE) mesh at a resolution of 19 by 51 elements in the xy plane. Using a conventional pre-calibrated material model one can compute woven composites responses at each material point for the macroscale model; however, ad-hoc nature of such models limit predictability. 

In this work, instead of using a materials law for the woven structural analysis, woven RVEs are deployed at each material points which compute woven material responses on the fly. An important issue that needs to be addressed in the concurrent modeling of the woven composite is the yarn material properties. As the yarn phase is essentially a mixture of long fibers and epoxy matrix, it can be assumed to behave similarly to the UD composite. With this assumption, the yarn behavior can be computed using UD RVEs. Coupling the macroscale model, mesoscale woven RVEs, and microscale UD RVEs, one arrives at a 3-scale multiscale model for woven structures. This 3-scale model requires one to resolve several woven and UD RVEs. Such 3-scale concurrent model can be computationally prohibitive considering RVE models usually contain millions of integration point. Therefore, in the current work, we proposed a 3-scale concurrent multiscale modeling approach with a reduced order modeling method that alleviates drastic computational cost brought by RVEs. Under this 3-scale concurrent scheme, responses of all RVEs are evaluated during the deformation process of the  tensile sample. For the modeling purpose, first we construct ROMs for the woven and UD RVE, and later these precomputed ROMs are used for concurrent analsysi of the multiscale model. Note unless otherwise clarified, all microstructure RVEs are built for cured composite materials. Carbon fiber and epoxy matrix are considered for the materials and the properties are taken from Table Table \ref{table:woven_material_const}. The epoxy matrix is assumed to have elasto-plastic behavior modeled with isotropic hardening following the yielding curve $\bar{\sigma}=30+50\bar{\epsilon}^{0.15}$, stress unit in MPa. 

In the 3-scale concurrent model, the mesoscale model provides material responses in each macroscale material point. In a similar fashion, the microscale model provides material responses in each mesoscale yarn material point. Here, the mesoscale model and the microscale model are identified as the woven RVE and the UD RVE, respectively. Modeling of two RVEs lays down the foundation of 3-scale concurrent modeling.

\begin{figure}[h]
	\centering
	\begin{subfigure}[b]{.4\linewidth}
	   		\centering
		\includegraphics[width=\linewidth]{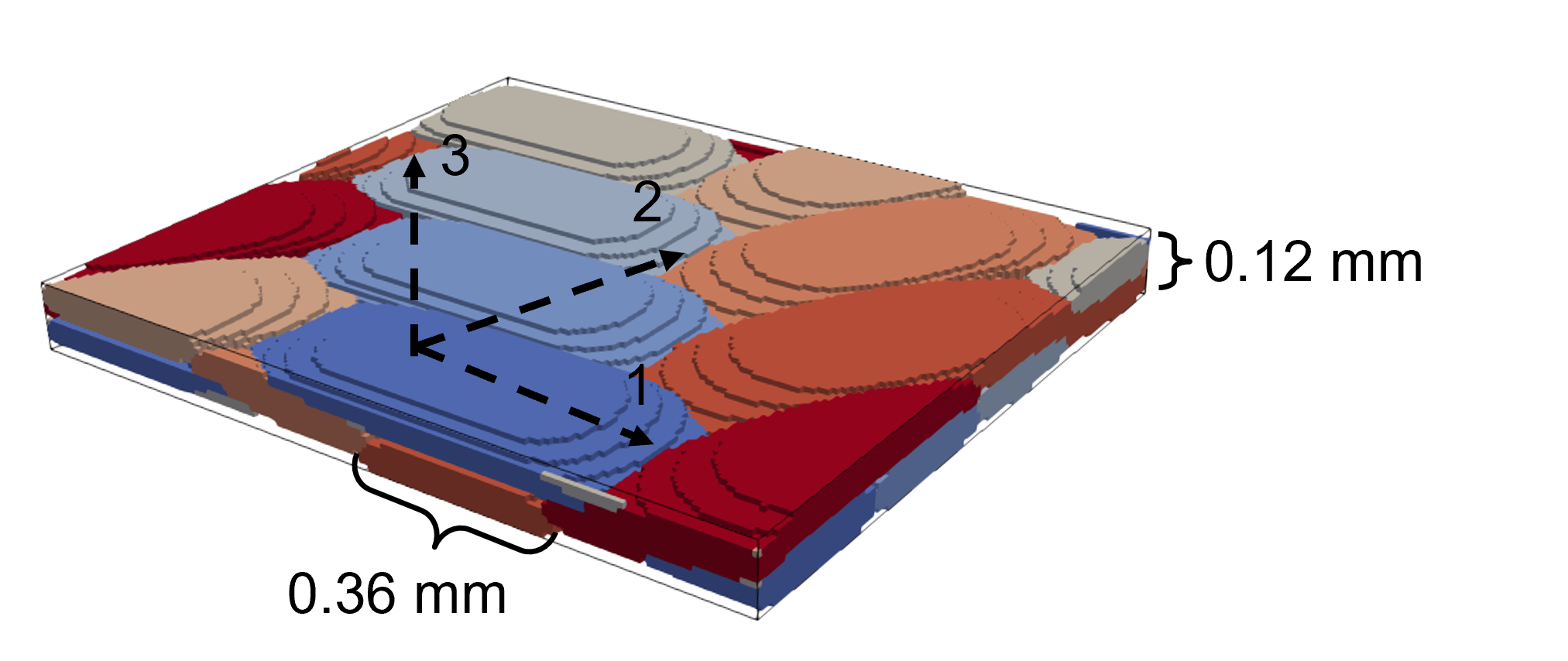}
		\caption{Mesoscale woven RVE} 
	\end{subfigure}
	
		\begin{subfigure}[b]{.4\linewidth}
	   		\centering
		\includegraphics[width=\linewidth]{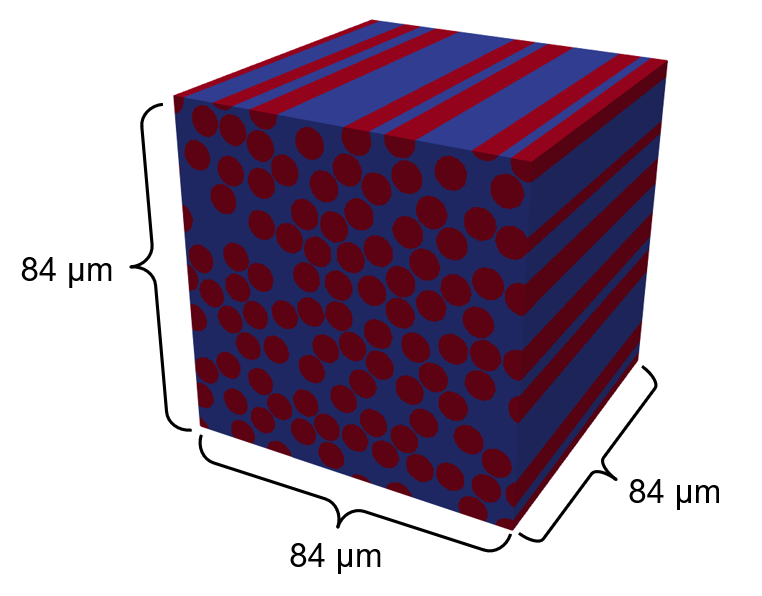}
		\caption{Microscale UD RVE} 
	\end{subfigure}
\caption{ (a) Mesoscale woven RVE with 60$^{\circ}$ yarn angle. The matrix phase is hidden for clarity so yarns can be observed.UD RVE with 44$\%$ fiber volume fraction. The dimension of the RVE is 84 $\mu$m by 84 $\mu$m by 84 $\mu$m.}	
	\label{fig:RVEs}
\end{figure}

The woven RVE model is constructed in TexGen \cite{long2011modelling}. As shown in Figure \ref{fig:RVEs} (a), the yarn width is defined as 0.36 mm. Four weft and four warp yarns are considered in the woven RVE model. The warp yarns are rotated in the 12 plane for 30$^{\circ}$ in the clockwise direction to setup the 60$^{\circ}$ yarn angle (denoted by 1 and 2'). The overall thickness of the RVE is 0.12 mm, with a yarn volume fraction of 85 $\%$. The woven RVE contains 149 voxels along 1 direction, 129 voxels along 2 direction, and 13 voxels along 3 direction. The woven RVE can be used to compute woven material behavior, once the basic constitutive laws are integrated. Here, the epoxy material is assumed to be isotropic with J2 plasticity following material properties given in Section \ref{sec:3pt-3-scale} and hardening properties above. For the yarn phase, the microscale UD RVE model is needed in order to compute its constitutive law. 

The UD RVE is generated following the methodology described in \cite{gao2020predictive}. The methodology utilizes a Monte Carlo approach to pack a square domain with circles, forming the cross section of the UD RVE. The fiber volume fraction needs to be given in order to perform the packing procedure. However, in the woven sample preparation process, the overall carbon fiber volume fraction is measured to be between 44$\%$ and 48$\%$. This means one needs to back-calculate the UD fiber volume fraction using the aforementioned numbers. Note that the overall fiber volume fraction is simply defined as the product of (fiber volume fraction in UD RVE)$\times$(yarn volume fraction in woven RVE), two UD RVEs are modeled with 51$\%$ and 56$\%$ fiber volume fractions. The UD RVEs are shown in Figure \ref{fig:RVEs}(b). Each UD RVE contains 360$^3$ voxels. 

\begin{figure}
	\centering
	\includegraphics[width=0.8\textwidth]{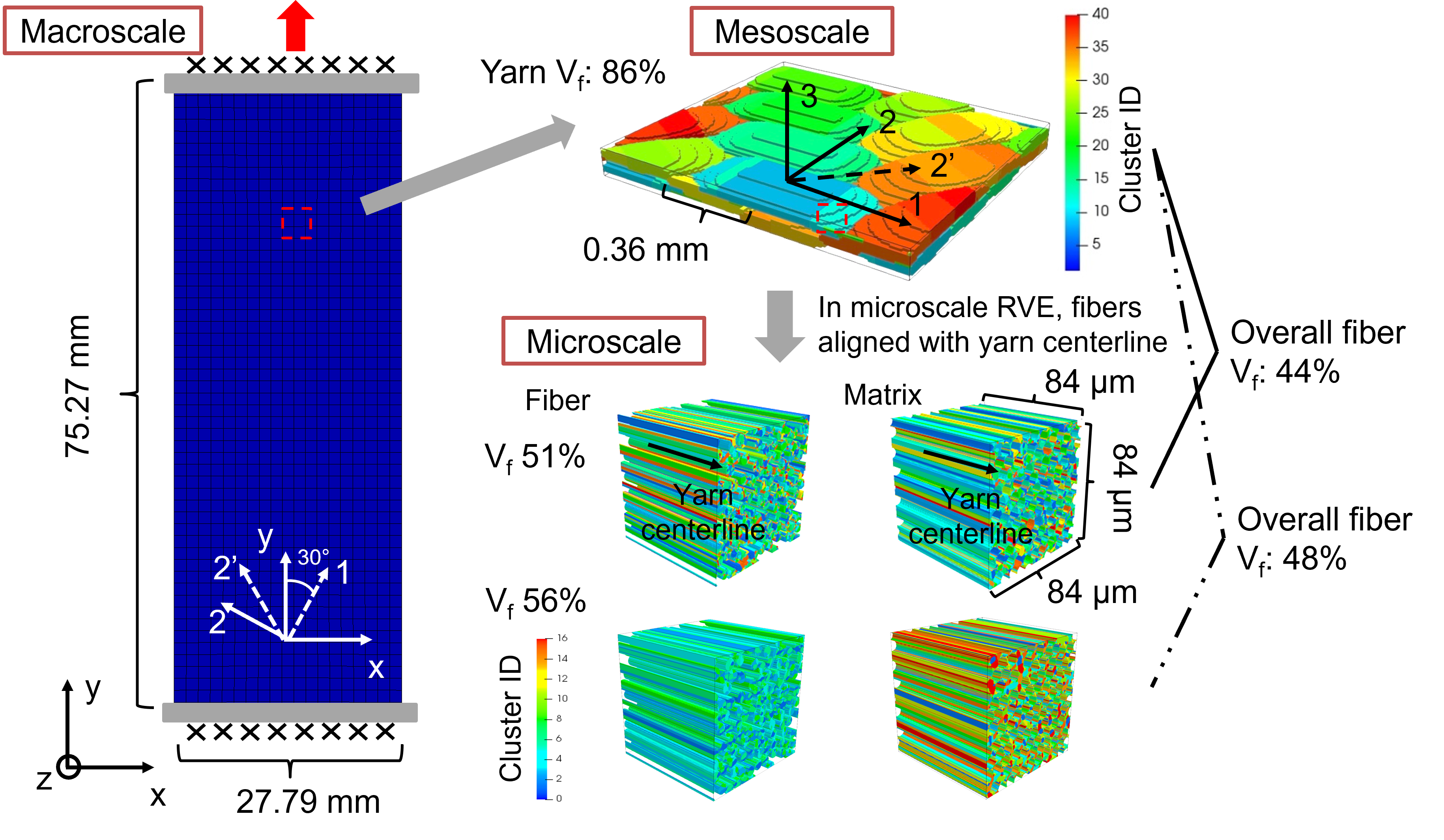}
	\caption{3-scale concurrent multiscale model for the cured 60$^\circ$ woven tensile test. The cluster distributions for mesoscale woven ROM of the woven RVE (yarn phase) and microscale UD ROM of the UD RVE(fiber and matrix phase) are depicted. Two different UD RVEs are used to capture sample fiber volume fraction upper and lower bounds.}
	\label{fig:b_e_3-scale_setup}
\end{figure}

As mentioned earlier, the 3-scale multiscale modeling is deployed for modeling the cured 60$^\circ$ woven composite. The overall 3-scale modeling framework is shown in Figure \ref{fig:b_e_3-scale_setup}, where the macroscopic material responses are provided by woven RVE with yarn angle of 60$^\circ$. In Figure \ref{fig:b_e_3-scale_setup}, the macroscale test sample is modeled as a Finite Element (FE) model with the same geometric specifications as the real coupon. The upper side of the sample is pulled upwards as  suggested by the red arrow, and the bottom side is fixed in all directions. The FE model is constructed in LS-DYNA \cite{manual2007version}, and contains 19 elements in the x direction, 53 elements in the y direction, and 1 element in the z direction. The model is meshed with solid elements, and each element contains one integration point. At each integration point, its material responses are computed using a 60$^\circ$ woven RVE in the 3-scale concurrent multiscale modeling framework, as illustrated.
In addition, the overall fiber volume fraction measured for the physical test sample shown in Figure \ref{fig:b_e_3-scale_setup} is between 44$\%$ and 48$\%$. The overall fiber volume fraction is defined as the product of yarn volume fraction and UD fiber volume fraction. As the range for the fiber volume fraction of the sample is given, the 3-scale model should be constructed following that range. This means two woven tensile 3-scale models are generated: one with overall fiber volume fraction of 44$\%$ and one with fiber volume fraction of 48$\%$. This is done through fixing the yarn volume fraction, and assigning two different UD RVEs: one with 51$\%$ fiber volume fraction and one with 56$\%$ fiber volume fraction. Those two cured woven tensile test models are expected to predict the lower bound and upper bound of the force vs. axial strain curves, forming a band that encompasses all experimental data.

%% file: result.tex
\section{Results and discussion}
\label{sec:result}
\begin{figure}[!htb]
	\centering
	\includegraphics[width=0.7\textwidth]{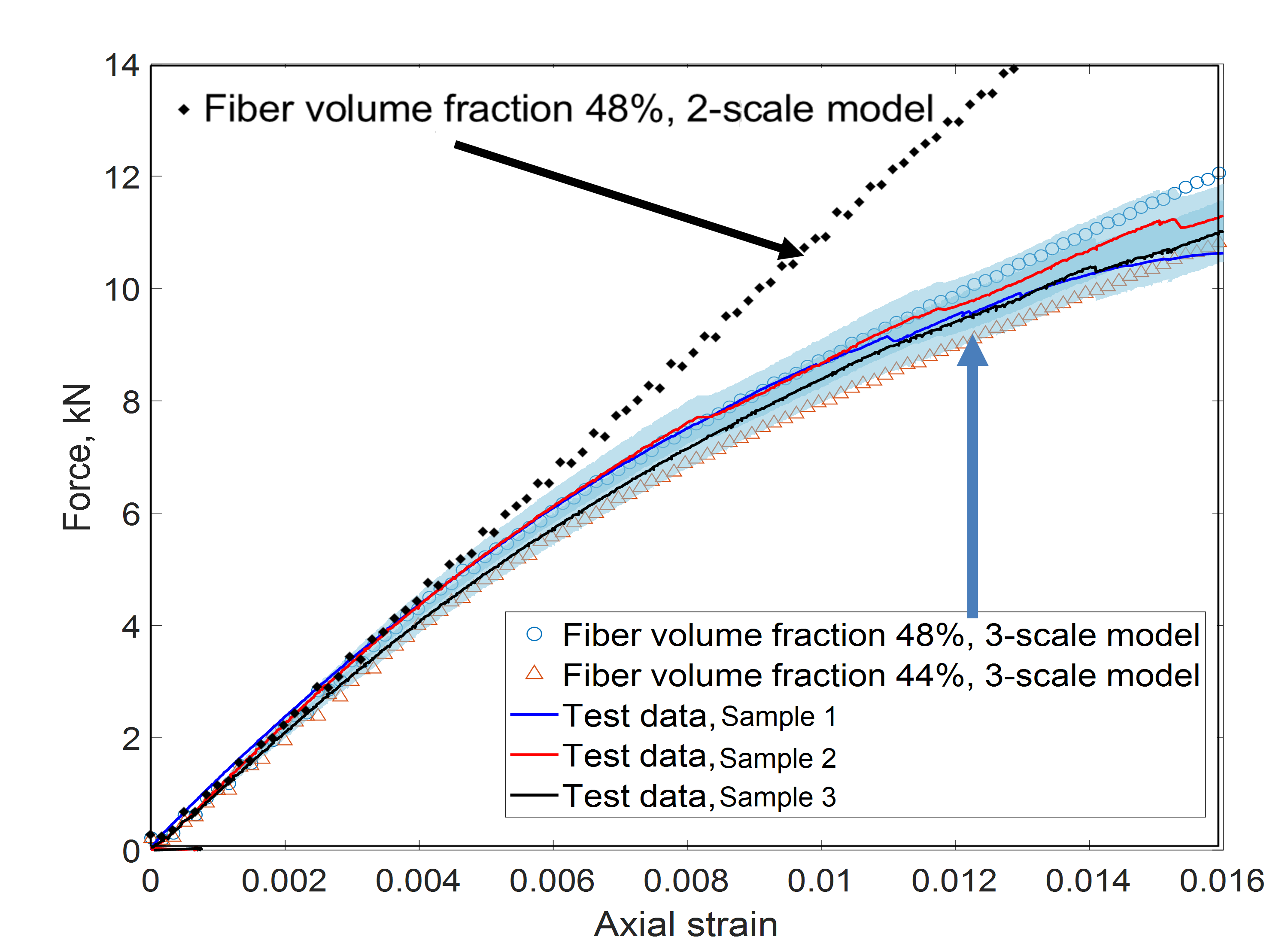}
	\caption{Loading force vs. axial strain plots for 2-scale and 3-scale concurrent multiscale predictions and experimental data. The sample fails after 0.016 axial strain.}
	\label{fig:b_e_2-scale_results}
\end{figure}

As described in the previous section a three-scale concurrent multiscale model is set up for the non-orthogonal woven bias extension sample. From the concurrent multiscale analysis in LS-DYNA, the loading force along the y direction is outputted for the entire loading process. The axial strain is computed for the sample so the force vs. axial plot can be generated and compared with the experimental data. The numerical prediction is plotted along with the experimental data obtained for three samples, identified as samples 1, 2, and 3. Those three samples all have the same geometry and are used to ensure the repeatability of the tensile test. As shown in Figure \ref{fig:b_e_2-scale_results}, three force vs. axial strain curves are plotted for three tensile test samples and they all have the same trend. Three curves from the samples are very close to each other, suggesting a good repeatability of the cured woven bias-extension test. From the figure it is clear that considering UD microstructure to approximate yarn responses is necessary. At this point, it has been shown that a UD RVE with only elastic material laws for the fiber and matrix can be used to compute the effective stiffness of the UD RVE. In the 3-scale model, UD RVE has a matrix phase with elasto-plastic material behavior. Assuming the matrix phase in the UD RVE behaves elastically, the 3-scale model automatically degenerates to a 2-scale model. The tensile sample with 48$\%$ fiber volume fraction is performed using both the 2-scale and 3-scale model, and the force vs. axial strain curve is also shown in Figure \ref{fig:b_e_2-scale_results}.

The 3-scale prediction agrees well and captures the trend of the experimental data. However, the results from the 2-scale model, which ignores matrix plasticity, diverges drastically from the test data when the axial strain is larger than 0.004. The flat curve suggests that the 2-scale model behaves elastically throughout the entire virtual tensile test, and it fails to correctly predict the woven tensile sample responses. The results suggest the matrix plasticity in the UD RVE matrix phase has a significant contribution to the nonlinear behavior of the woven tensile sample. Therefore, the 60 $^\circ$ woven tensile sample has to be constructed using a 3-scale model so that the material responses can be captured using detailed microstructure. As mentioned before, two cured bias extension models are built in order to examine the effect of overall fiber volume fraction in the sample.The predictions made by 3-scale model with 44$\%$ and 48$\%$ fiber volume fraction, as well as three sets of experimental data, are visualized in Figure \ref{fig:b_e_2-scale_results} . Both predictions agree well with the test data. To be specific, the sample with 48$\%$ captures draws the upper bound for all test data, and the sample with 44$\%$ draws the lower bound. Two 3-scale predictions from our 3-scale concurrent model successfully draw a band that captures all three experimental data curves. This means the present methodology can potentially replace partial experimental trial, and provide a high quality prediction of the woven composite responses (in this case, the tensile responses of the 60$^\circ$ woven). Note that the entire virtual testing process requires zero calibration effort, unlike the traditional phenomenological material law approach.

\begin{figure}
	\centering
	\includegraphics[width=0.85\textwidth]{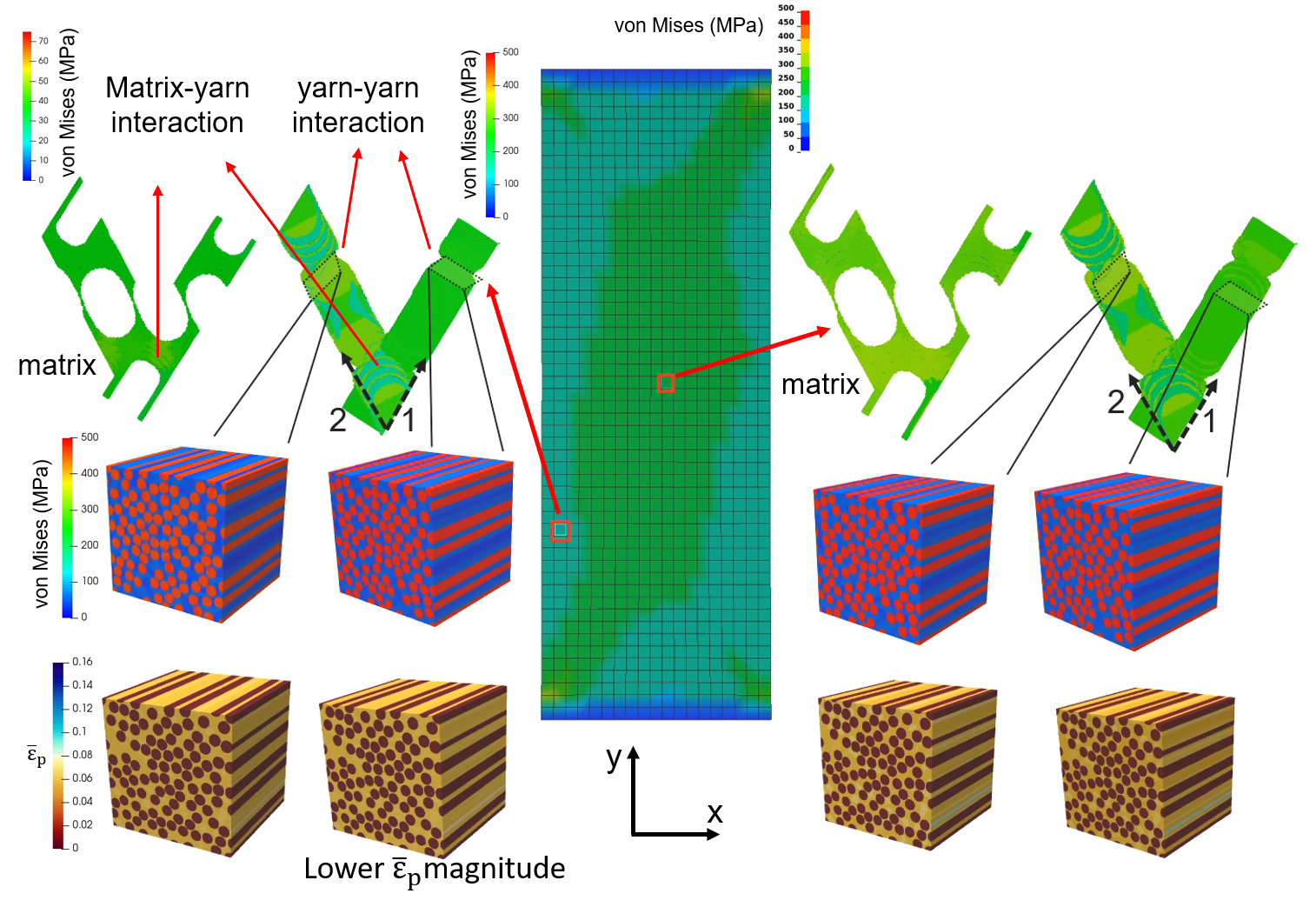}
	\caption{Visualization at axial strain of 0.006: Cured woven bias-extension sample von Mises stress contour, mesoscale yarn and matrix von Mises stress contour, and microscale von Mises stress and effective plastic strain contours.}
	\label{fig:woven_3_eps0p006}
\end{figure}

\begin{figure}
	\centering
	\includegraphics[width=0.85\textwidth]{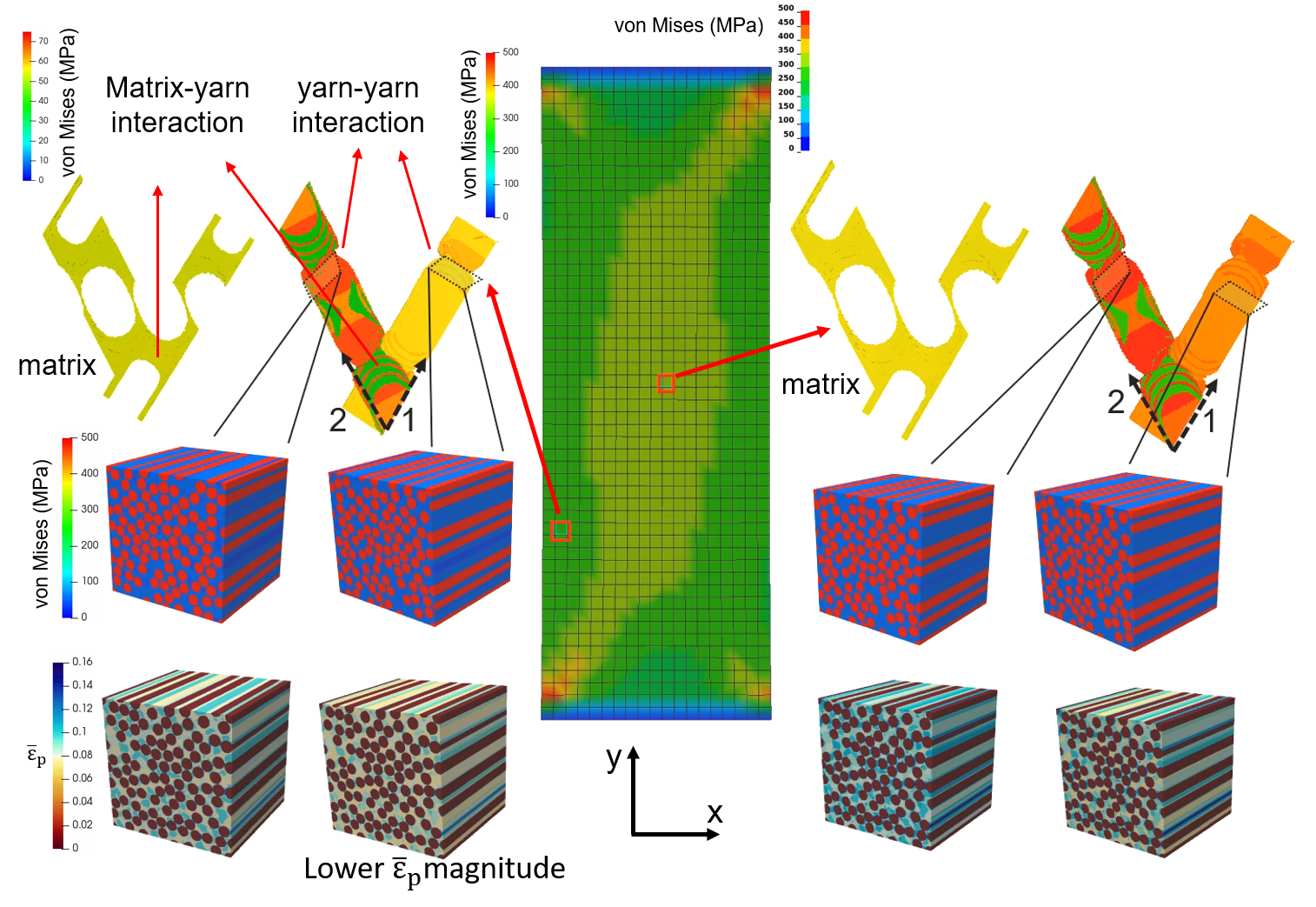}
	\caption{Visualization at axial strain of 0.01: Cured woven bias-extension sample von Mises stress contour, mesoscale yarn and matrix von Mises stress contour, and microscale von Mises stress and effective plastic strain contours.}
	\label{fig:woven_3_eps0p01}
\end{figure}

\begin{figure}
	\centering
	\includegraphics[width=0.85\textwidth]{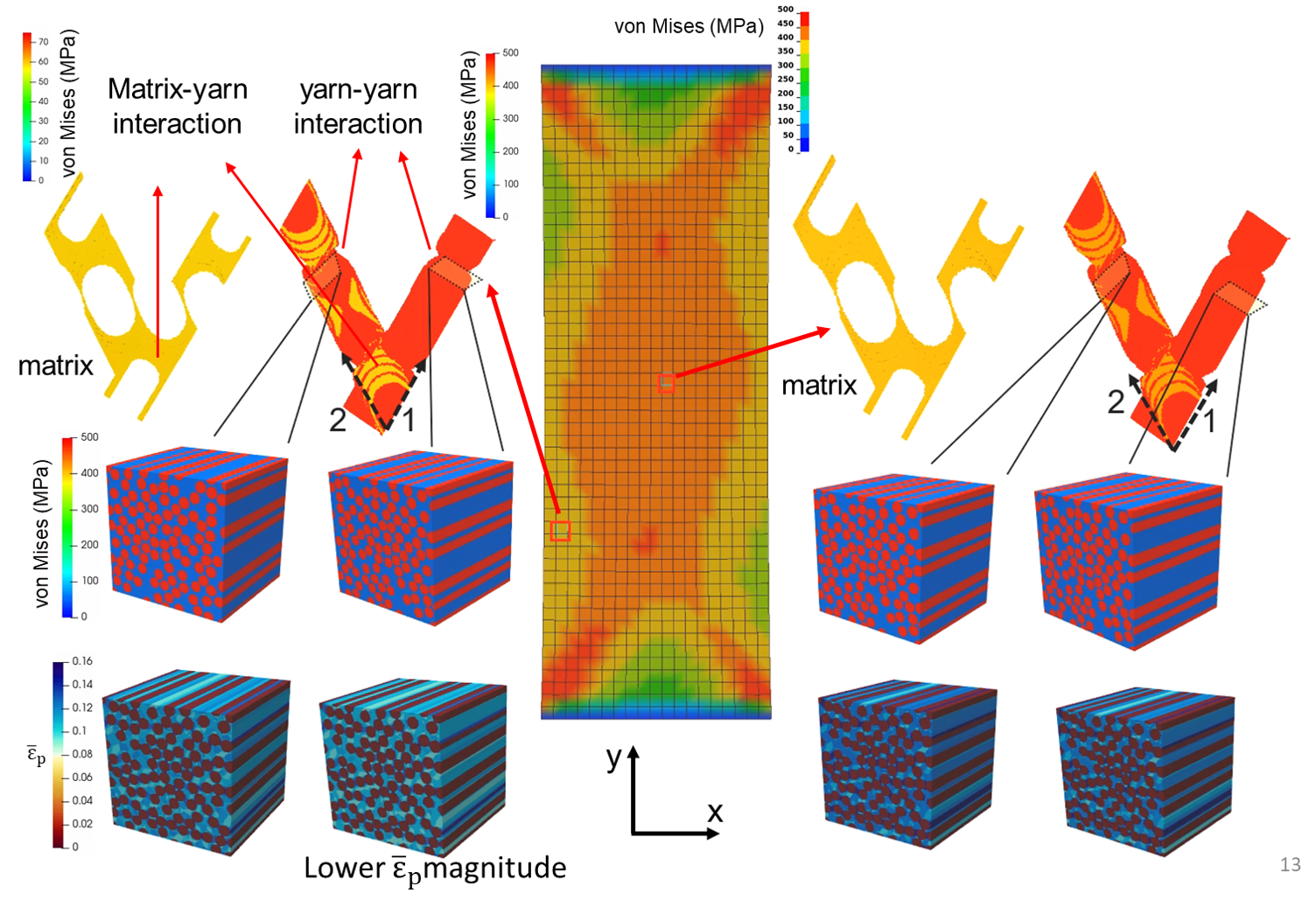}
	\caption{Visualization at axial strain of 0.016: Cured woven bias-extension sample von Mises stress contour, mesoscale yarn and matrix von Mises stress contour, and microscale von Mises stress and effective plastic strain contours.}
	\label{fig:woven_3_eps0p016}
\end{figure}

In addition, the deformed cured woven bias-extension sample and the local yarn-UD plastic strain distribution are visualized in Figure \ref{fig:woven_3_eps0p006}, Figure \ref{fig:woven_3_eps0p01}, and Figure \ref{fig:woven_3_eps0p016}  at the axial strain of 0.006, 0.01, and 0.016, respectively. In each figure, mesoscale woven RVEs at two boxed locations are visualized. For both woven RVEs, part of the woven RVE, including one weft and one warp yarns, is visualized in order to examine the local shear deformation. On each yarn, one integration point is picked out and its underneath UD RVE is visualized to examine the yarn plastic strain accumulation through out the entire deformation process.

In Figure \ref{fig:woven_3_eps0p006}, the axial strain has reached 0.006 and one can see stress starts to concentrate in the middle due to 60$^{\circ}$ yarn angle. There is a X shape stress band in the sample, showing higher stress magnitude than the rest of the sample.
The woven RVE in the stress concentration region is shown on the right of Figure \ref{fig:woven_3_eps0p006}; its yarn and matrix phase both demonstrate higher stress magnitude than those from the woven RVE, shown on the left, outside the stress concentration region. Moreover, taking the cross sections of both yarns on two woven RVEs, microscale UD stress and effective plastic strain ($\bar{\varepsilon}_p$) evolution are revealed. The stress magnitude of UD RVEs on the right is higher than RVEs on the left, showing a consistent trend observed on the woven RVEs. The $\bar{\varepsilon}_p$ contour shows the UD goes into plastic deformation, which is the source of the nonlinearity observed in Figure \ref{fig:b_e_2-scale_results}. In addition, the $\bar{\varepsilon}_p$ contours on the right show higher magnitude than contours on the left. 

When the axial strain reaches 0.01, Figure \ref{fig:woven_3_eps0p01} shows detailed evolution of those physical fields introduced in the previous figure. Here, on the right the stress contours of two UD RVEs are very close to each other, but the $\bar{\varepsilon}_p$ contours suggest UD in the yarn along 2 direction suffer larger deformation due to higher $\bar{\varepsilon}_p$ magnitude. One the left, one can see two UD RVEs have less $\bar{\varepsilon}_p$ than those on the right. The difference in the stress contours for yarns along the 2 direction is less obvious. This suggests that the capture of microscale $\bar{\varepsilon}_p$ can play an important role in governing yarn failure, providing an alternative to the stress-based failure criteria in the multiscale models.

Finally, as the axial strain reaches 0.016, Figure \ref{fig:woven_3_eps0p016} shows that the difference in woven RVE stress contours is almost negligible. However, the contrast in UD RVE $\bar{\varepsilon}_p$ can be used to quantify the difference. Most of the matrix phase on the right reaches $\bar{\varepsilon}_p$ of 0.16, while the matrix phase on the left shows $\bar{\varepsilon}_p$ is around 0.12, suggesting a 33$\%$ difference. A movie for the concurrent evolution of the stress and strain field in the cured woven bias-extension sample and the RVE's at different scales is provided as Supplementary material (S2).
 
 In terms of the computational time, the study case shown here was performed on a High Perform Cluster over two nodes, each node has 28 processors. The entire computation takes 40 hours, and the physics over 969 woven RVEs (one in each macroscale integration points) and 15504 UD RVEs (one in each yarn clusters) are revealed simultaneously. Comparing to a model with full voxel mesh in woven and UD RVEs, the speedup enabled by ROM is estimated to be on the order of 10\textsuperscript{8}.
 
 The 3-scale bias extension model successfully reproduced the nonlinear stress and strain curve observed in the experiment. The detailed physics revealed by the 3-scale model provides convenience in analyzing nonlinear behavior of the woven composite structure. Such details provided by the 3-scale model make it possible to capture local matrix plastic strain concentration in determining matrix failure. This concurrent multiscale tool can be useful for the design and analysis of different designs of the architectured composites such as woven and braided structures.